# Room-temperature correlated states in twisted bilayer MoS$_2$

Fanfan Wu[1,2,⊥], Qiaoling Xu[3,4,⊥], Qinqin Wang[1,2], Yanbang Chu[1,2], Lu Li[1,2], Jian Tang[1,2], Jieying Liu[1,2], Jinpeng Tian[1,2], Yiru Ji[1,2], Le Liu[1,2], Yalong Yuan[1,2], Zhiheng Huang[1,2], Jiaojiao Zhao[1,2], Xiaozhou Zan[1,2], Kenji Watanabe[5], Takashi Taniguchi[6], Dongxia Shi[1,2,3], Gangxu Gu[1,2], Yang Xu[1,2], Lede Xian[3]*, Wei Yang[1,2,3]*, Luojun Du[1,2]*, Guangyu Zhang[1,2,3]*

[1]*Beijing National Laboratory for Condensed Matter Physics, Institute of Physics, Chinese Academy of Sciences, Beijing 100190, China*
[2]*School of Physical Sciences, University of Chinese Academy of Sciences, Beijing 100049, China*
[3]*Songshan Lake Materials Laboratory, Dongguan, Guangdong 523808, China*
[4]*College of Physics and Electronic Engineering, Center for Computational Sciences, Sichuan Normal University, Chengdu 610068, China*
[5]*Research Center for Functional Materials, National Institute for Materials Science, 1-1 Namiki, Tsukuba 305-0044, Japan*
[6]*International Center for Materials Nanoarchitectonics, National Institute for Materials Science, 1-1 Namiki, Tsukuba 305-0044, Japan*
*Email: luojun.du@iphy.ac.cn; xianlede@sslab.org.cn; wei.yang@iphy.ac.cn; gyzhang@iphy.ac.cn*

**Moiré superlattices have emerged as an exciting condensed-matter quantum simulator for exploring the exotic physics of strong electronic correlations. Notable progress has been witnessed, but such correlated states are achievable usually at low temperatures. Here, we report the transport evidences of room-temperature correlated electronic states and layer-hybridized SU(4) Hubbard model simulator in AB-stacked MoS$_2$ homo-bilayer moiré superlattices. Correlated insulating states at moiré band filling factors $v$ = 1, 2, 3 are unambiguously established in twisted bilayer MoS$_2$. Remarkably, the correlated electronic states can persist up to a record-high critical temperature of over 285 K. The realization of room-temperature correlated states in twisted bilayer MoS$_2$ can be understood as the cooperation effects of the stacking-specific atomic reconstruction and the resonantly enhanced interlayer hybridization, which largely amplify the moiré superlattice effects on electronic correlations. Furthermore, extreme large non-linear Hall responses up to room-temperature are uncovered near correlated insulating states, demonstrating the quantum geometry of moiré flat conduction band.**

**Introduction**

Moiré superlattices formed by vertically stacking atomic layers with a twist and/or a lattice mismatch, introduce a new length and energy scale to engineer the electronic band structure and represent a highly controllable quantum platform for exploring a wide variety of emerging correlated physics[1-3]. One well-known manifestation is the magic-angle twisted bilayer graphene, where the formation of moiré flat bands strongly enhances the ratio $U/W$ by significantly suppressing the $W$ ($U$ and $W$ are the electron-electron interaction and electronic bandwidth, respectively) and hence results in a rich phase diagram of correlated electronic states, such as Mott insulating states[4], Chern insulator states[5], orbital ferromagnetism[6,7] and unconventional superconductivity[6,8]. However, because the electron-electron interaction $U$ is considerable small (~10 meV)[4], these correlated states in twisted graphene system can survive only at temperatures of a few K, limiting their tunability in experiments and hindering their applications in quantum technology. Therefore, it is particularly important to explore additional moiré superlattices with large electron-electron interaction $U$, which may enable the possibilities of correlated electronic states at high temperatures.

Given that the electron-electron interaction $U$ is inversely proportional to the dielectric constant, transition metal dichalcogenide (TMD) moiré superlattices with relatively small dielectric constants represent a potential platform to realize high-temperature correlated physics[2,9]. Indeed, a host of exotic correlated electronic states, such as generalized Wigner crystal states, excitonic insulator, quantum criticality and quantum anomalous Hall effect have been uncovered in TMD homo- and hetero-structure superlattices at much higher critical temperatures than in graphene moiré superlattices[10-17]. Notably, correlated insulating phases up to about 150 K have been realized at half-filling of the first moiré miniband in angle-aligned $WSe_2/WS_2$ and $MoSe_2/WS_2$ hetero-bilayer moiré superlattices[18,19]. The critical temperatures of correlated electronic states in these TMD heterostructure superlattices, although significantly increased compared to that in graphene moiré superlattices, are still well below room-temperature.

In this work, we fabricate high-quality, encapsulated, dual-gated $MoS_2$ homo-bilayer moiré superlattice devices and report the transport evidences of room-temperature correlated electronic states in AB stacked samples with a twist angle of ~57.5°. Insulating states at doping density of one, two and three electrons per moiré superlattice site are uncovered unambiguously in twisted bilayer $MoS_2$. Together with first-principle calculations, we uncover that these insulating states correspond to correlated electronic phases at moiré band filling factors $v = 1, 2, 3$, evidencing a layer-hybridized SU(4) correlated Hubbard model simulator. This is in sharp contrast to previous TMD moiré superlattice results of single-band SU(2) Hubbard model that correlation-induced Mott insulating phases can appear only at half-filling of moiré band with one hole/electron per superlattice site. Notably, the $v = 1$ correlated state can persist up to a record-high temperature of over 285 K and host a record-large Mott-Hubbard gap, demonstrating the ultra-strong electron correlations in $MoS_2$ homo-bilayer moiré superlattice. Furthermore, large non-linear Hall responses up to room-temperature are uncovered near the correlated insulating states, demonstrating the quantum geometrical properties of electron wavefunctions.

**Correlated electronic states in twisted bilayer MoS₂**

It is well known that the direct electrical-transport detection of the correlated electronic states is challenging in TMD semiconductor moiré superlattices[11,20]. To construct high-quality moiré superlattice devices for transport measurements, we adopt monolayer MoS$_2$ single crystal samples grown by van der Waals epitaxial technique, which show better electronic qualities than the exfoliated ones (Supplementary Fig. 1)[21,22]. Then hexagonal boron nitride (*h*-BN) encapsulated, dual-gated MoS$_2$ homo-bilayer moiré superlattice devices with multiple electrodes are fabricated utilizing the 'cut and stack' technique (see details in the Methods section)[23,24]. Figure 1a shows the schematic diagram of the twisted bilayer MoS$_2$ dual-gate devices. The twist angles are controlled to be at ~57.5° (i.e., AB stacking) or ~2.5° (i.e., AA stacking), producing a moiré superlattice with a periodicity close to 7-8 nm, as illustrated in Fig. 1b. The long periodic moiré potential would fold the band into the mini-Brillouin zone (Fig. 1c). Few-layer graphene less than 1 nm is used as the contact to MoS$_2$, which can endow good ohmic contact and facilitate the cryogenic transport[25]. Moreover, the dual-gate configuration enables us to independently tune the carrier density *n* and out-of-plane displacement field **D**. Here $n = (C_b V_b + C_t V_t)/e$ and $\boldsymbol{D} = (C_b V_b - C_t V_t)/2\varepsilon_0$, where *e* is the elementary charge, $\varepsilon_0$ denotes the vacuum permittivity, $C_b$ ($V_b$) and $C_t$ ($V_t$) are the geometrical capacitances per area (applied voltages) for bottom and top gates, respectively.

Figure 1e shows the colour plot of four-terminal longitudinal resistance as a function of electron densities at various temperatures (5 K–60 K) for 57.15° twisted bilayer MoS$_2$ (Device #1, the corresponding optical image can be found in Fig. 1d). Unless otherwise specified, all results in the main text are taken from Device #1. Remarkably, apart from the diverging resistance peaks at *n* = 0 that mark the intrinsic band edge of MoS$_2$, four strong resistance peaks at $n = n_0, 2n_0, 3n_0, 4n_0$ are clearly observed. $n_0 = 2.85 \times 10^{12}$ cm$^{-2}$ is equal to 1/A and thus corresponds to the moiré density that one electron per superlattice site, where $A = \frac{\sqrt{3}}{8} \frac{a^2}{[\sin(\frac{60-\theta}{2})]^2}$ denotes the moiré unit cell area with $a = 0.315$ nm the MoS$_2$ lattice constant and $\theta$ is the twist angle. In other words, the four strong resistance peaks at $n = n_0, 2n_0, 3n_0, 4n_0$ coincide with one, two, three, four electrons per moiré superlattice site—that is, moiré band filling factors *v* = 1, 2, 3, 4. It is noteworthy that the resistance corresponding to *v* = 1 is too large to be detected at temperatures below 28 K. Figure 1f shows a plot of the resistance versus electron-band carrier densities for five different twisted bilayer MoS$_2$ devices. All the five devices exhibit apparent resistance peaks at moiré band filling factors *v* = 1, 2, 3. The weak insulating state at *v* = 4 is because that the measuring temperature is above 20 K.

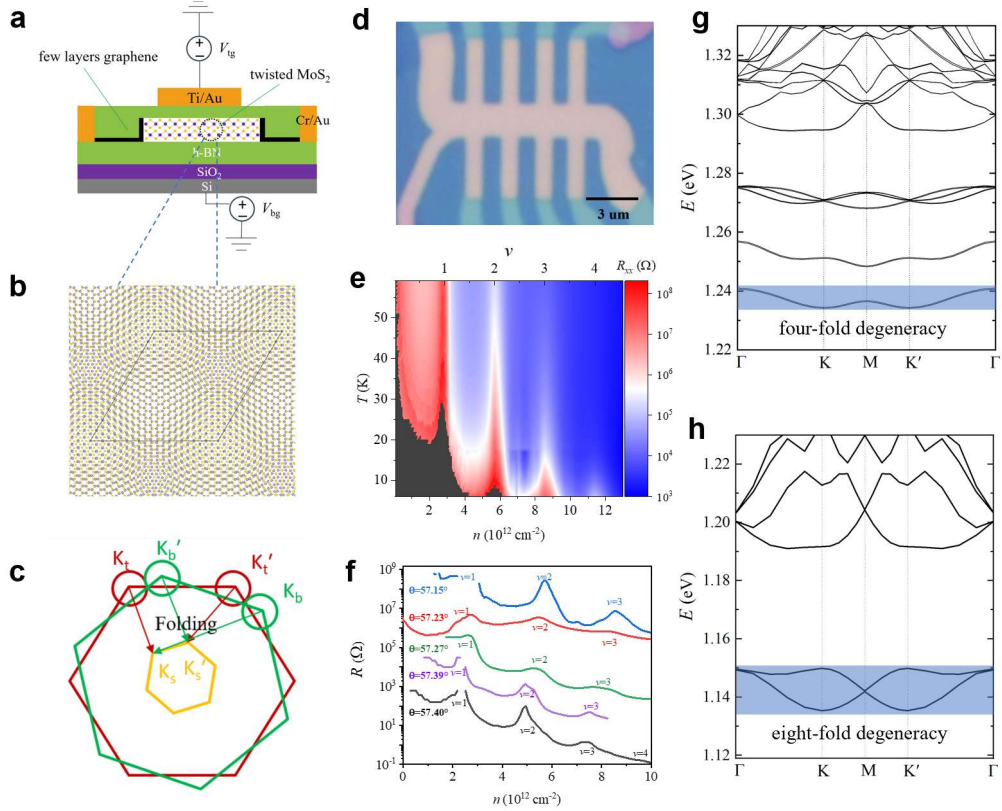

**Fig. 1 | Layer-hybridized SU(4) correlated Hubbard model in twisted bilayer MoS$_2$. a,** Schematic geometry of the twisted bilayer MoS$_2$ moiré superlattice device, encapsulated in *h*-BN with thicknesses of about 20–30 nm. **b,** Illustration of the moiré superlattice formed by ~57.5° twisted bilayer MoS$_2$. **c,** Schematic illustration of moiré mini-Brillion zone. **d,** Optical image of 57.15° twisted bilayer MoS$_2$ device #1. **e,** Colour plot of the four terminal longitudinal resistance against the carrier densities and temperatures of device #1, showing clearly four prominent insulating states at moiré band filling factors $v$ = 1, 2, 3 and 4. **f,** Resistance as a function of carrier densities for five different twisted bilayer MoS$_2$ devices. The curves are offset for clarity. **g-h,** The single-particle energy dispersion of 56.85° twisted bilayer MoS$_2$ in the first mini-Brillouin zone with (**g**) and without (**h**) consideration of the layer hybridization. The first electron moiré flat miniband is outlined by blue shadow.

To gain more insight of these insulating states, we thus perform density functional theory (DFT) calculations on the commensurate 56.85° twisted bilayer MoS$_2$. The calculated band structure with considering the layer hybridization is shown in Fig. 1g (see detailed information in Supplementary Note 2). Clearly, ultra-flat moiré minibands emerge at the conduction band bottom with a band width less than 10 meV (Supplementary Fig. 2). The flat moiré minibands are contributed from the original ±K valley states in the two constituent MoS$_2$ layers as confirmed by their charge density distribution (Supplementary Fig. 4). Because of negligible spin-orbit coupling (only ~3 meV)[26], four spin-degenerate states from the ±K valleys of the two constituent layers are folded to moiré mini-Brillion zone (Fig. 1c). With the atomic reconstruction in the moiré scale, eight-fold degenerate flat minibands are formed (including two spin, two valley and two layer) when the layer hybridization is artificially turned off (Fig. 1h). Owing to the perfect band alignment, interlayer hybridization is resonantly enhanced in twisted bilayer MoS$_2$. After turning on the layer hybridization, the eight-fold degenerate flat minibands split into two sets of four-fold degenerate flat minibands as layer bonding and anti-boding orbitals (Fig. 1g and Supplementary Fig. 3). Consequently, the resistance peak at $v = 4$ can be well understood as the full filling of the first electron moiré flat miniband with a reduction in the density of states. By contrast, the insulating states at filling factors $v = 1, 2, 3$ correspond to the partial fillings of the first moiré flat conduction miniband and defy the description by single-particle band structure paradigm. This indicates that strong electron-electron interactions exist within moiré flat conduction minibands in the twisted bilayer MoS$_2$, which lift the flavour degeneracy, give rise to full spin-valley polarized flat bands and drive Mott-like insulating states with correlation-induced gaps at moiré band filling factors $v = 1, 2, 3$.

Interestingly, the presence of correlated insulator states at $v = 1, 2, 3$ suggests that the moiré conduction bands of twisted bilayer MoS$_2$ would emerge as a layer-hybridized SU(4) correlated Hubbard model simulator. This is in stark contrast to previous TMD moiré superlattice results described by the one-orbital SU(2) correlated Hubbard model (that is, correlation-induced insulating states at integer fillings appear only at half-filling of moiré flat band, i.e., $v = 1$)[13,17,18] and offer extraordinary opportunities to underpin new phenomena, such as orbitally selective Mott phase[27], SU(4) chiral spin liquid and exciton supersolid phases[28].

**Room-temperature correlated electronic states**

By comparing the band structures with (Fig. 1g) and without (Fig. 1h) considering the layer hybridization, electronic bandwidth $W$ is significantly reduced for the former. This indicates the resonantly enhanced interlayer hybridization induced by the perfect band alignment in twisted bilayer MoS$_2$, which can strongly amplify the moiré superlattice effects, and may offer the possibility of high-temperature correlated physics[29]. To confirm this, we thus perform temperature dependent transport measurements. Figure 2a presents the evolution of insulating states at $v = 1, 2, 3, 4$ as a function of temperature. For clarity, the smooth background has been subtracted (Supplementary Fig. 5). Strikingly, the correlated insulating states in twisted bilayer MoS$_2$ of AB stacking can persist up to high temperatures. For the $v = 1$ Mott-like insulating state, the critical temperature $T_c$ at which the insulating response onsets can reach a record-high value of over 285 K, much higher than those reported in magic-angle bilayer graphene (e.g., ~10 K)[6] and also better than the state-of-the-art results of TMD homo- and hetero-structure moiré superlattices (e.g., ~150 K)[18]. Similar values of $T_c$ are also confirmed by the $R$-$T$ curves (Supplementary Fig. 6). The ultra-high gap-closure temperatures implies that the electron-electron correlations are ultra-strong in MoS$_2$ homo-bilayer moiré superlattice. From the temperature-dependent transport behaviour, we can extract the activated gap sizes $\Delta$ of the correlated insulating states with a thermal-activation model $R \propto e^{-\Delta/(2k_BT)}$ ($k_B$, Boltzmann constant; $T$, temperature). Figure 2b gives the Arrhenius fits to resistance for the $v = 1$ Mott-like insulating state. Also see Supplementary Fig. 7 for the Arrhenius fits to resistance at $v$

= 2, 3 and 4. The derived correlated gaps at $v$ = 1, 2, 3, 4 are 126 meV, 73 meV, 30 meV and 5 meV, respectively (left panel of Fig. 2c). According to the correlated gaps, we illustrate the energy diagram of MoS$_2$ homo-bilayer moiré superlattice of AB stacking in right panel of Fig. 2c. The Hubbard bandwidth increases and the density of state decreases with energy due to interaction effects weaken with increasing $v$. It is noteworthy that both the gap-closure temperature of over 285 K and the Mott-Hubbard gap of 126 meV at $v$ = 1 are the highest values ever achieved among all the moiré superlattice systems (Fig. 2d)[4,6,11,15,17-19,24,30-34], confirming the giant electron correlations in MoS$_2$ homo-bilayer moiré superlattice with strong layer hybridization.

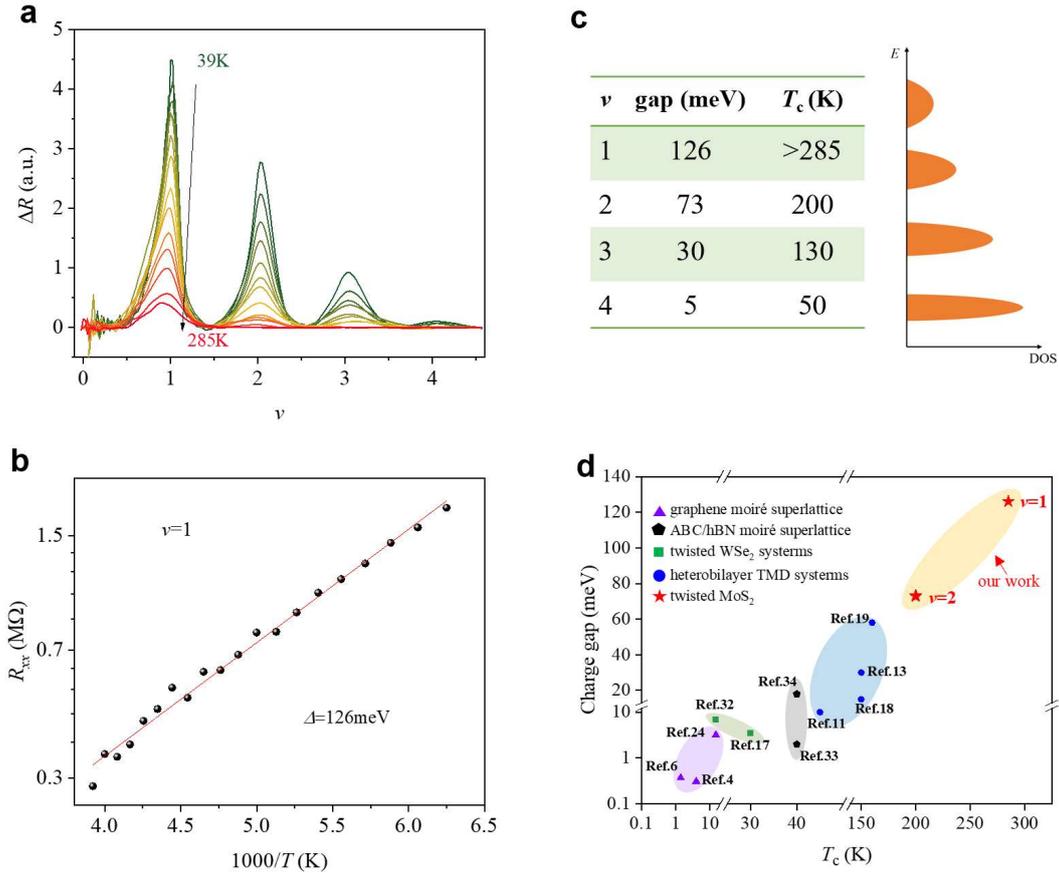

**Fig. 2 | Room-temperature correlated electronic states in MoS$_2$ homo-bilayer moiré superlattice. a,** Four-terminal longitudinal resistance plotted against filling factors at different temperatures from 39 K (green trace) to 285 K (red trace). The smooth resistance background has been subtracted. **b,** Plot of $R$ versus $1000/T$ for $v$ = 1 Mott-like insulating state. The extracted thermal excitation gap at $v$ = 1 through the thermal-activation function (red line) is about 126 meV. **c,** Left: the derived correlated gaps and gap-closure temperature $T_c$ at $v$ = 1, 2, 3 and 4. Right: schematic illustration of the density of states against energy for the four spin-valley polarized flat Hubbard bands driven by strong many-body interactions. **d,** Comparison of the correlated gaps and gap-closure temperatures of twisted bilayer MoS$_2$ with those in various kinds of moiré superlattices previously reported. Both the critical temperature (> 285 K) and the Mott-Hubbard gap (~126 meV) at $v$ = 1 are the largest value achieved so far among all the moiré superlattice systems.

**Stacking effects**

In contrast to the twisted bilayer graphene, the sublattice symmetry breaking in TMDs endows the distinct moiré superlattices for twist angles near 0° and 60°. Consequently, TMD moiré superlattices of AB and AA stacking orders are expected to exhibit divergent electronic band structures and electron-electron Coulomb interactions. To confirm this stacking order-governed moiré effect, we also perform the transport measurements on twisted bilayer $MoS_2$ with twist angles around 2.5° (four devices are tested). Interestingly, twisted bilayer $MoS_2$ of AA stacking do not show any correlation phenomena (Supplementary Fig. 8-9), in contrast to the case of AB stacked $MoS_2$ moiré superlattices. The absence of correlated insulating states in AA stacked $MoS_2$ moiré superlattices reveals the strong stacking effects on moiré physics and is consistent with the lack of flat conduction bands evidenced by our calculation (Supplementary Fig. 2) and previous multiscale theory[35].

**Non-linear Hall responses**

The ultra-strong electron correlations and room-temperature correlated insulating states, in principle, would enable high-temperature quantum phenomena in twisted bilayer $MoS_2$, such as second-order non-linear Hall effect intricately connected with the quantum geometrical property of Bloch wavefunctions (i.e., Berry curvature dipole)[36,37]. Figure 3a shows a schematic diagram for the non-linear Hall measurements. Figure 3b presents the non-linear Hall voltage $V^{2\omega}$ as a function of moiré filling factors for a 57° twisted bilayer $MoS_2$ device (Device #2). Here we show the non-linear Hall responses at a relatively high temperature of 60 K so that the $v = 1$ correlated insulating state can be well distinguished. Notably, strong non-linear Hall responses can be observed at filling factor near $v = 1$ and 2 correlated insulating states. When lowering the temperature (e.g., 30 K), strong non-linear Hall responses near filling factor of $v = 3$ can also be found (Supplementary Fig. 11). Figure 3c shows the $V^{2\omega}$ against the applied current $I^\omega$ for filling factor near $v = 1$ correlated insulating state at 60K. Second-harmonic Hall voltage $V^{2\omega}$ scales linearly with the square of $I^\omega$, confirming the nature of second-order non-linear Hall responses. The linear dependence of $V^{2\omega}$ on $(I^\omega)^2$ can also be found near filling factors of $v = 2$ and 3 (Supplementary Fig. 12).

Figure 3d shows the evolution of nonlinear Hall generation efficiency defined as $\eta = V^{2\omega}/(V^\omega)^2$ as a function of temperatures near filling factor of $v = 1$. At low temperature, the nonlinear Hall generation efficiency $\eta$ in twisted bilayer $MoS_2$ can reach larger than 360 V$^{-1}$, two orders of magnitude larger than those in bilayer and few-layer $WTe_2$[38,39]. Note that giant non-linear Hall responses have also be recently observed in twisted bilayer $WSe_2$ at half-filling[40]. With increasing temperature, $\eta$ gradually decreases. The coincidence of the non-linear Hall voltage maximum with the correlated insulating states, together with its temperature evolution, reveals that electron correlations play a crucial role in determining Berry curvature dipole and quantum geometry of flat conduction band wavefunction. Surprisingly, considerable non-linear Hall responses with $\eta = \sim 1$ V$^{-1}$ can even persist up to room-temperature for $v = 1$ moiré band filling (inset of Fig. 3d). This further confirms the room-temperature correlated insulating state and ultra-strong electron correlations in twisted bilayer MoS$_2$ of AB stacking. The sizable room-temperature non-linear Hall responses might be useful to realize technological advances in wireless radiofrequency rectification and frequency doubling.

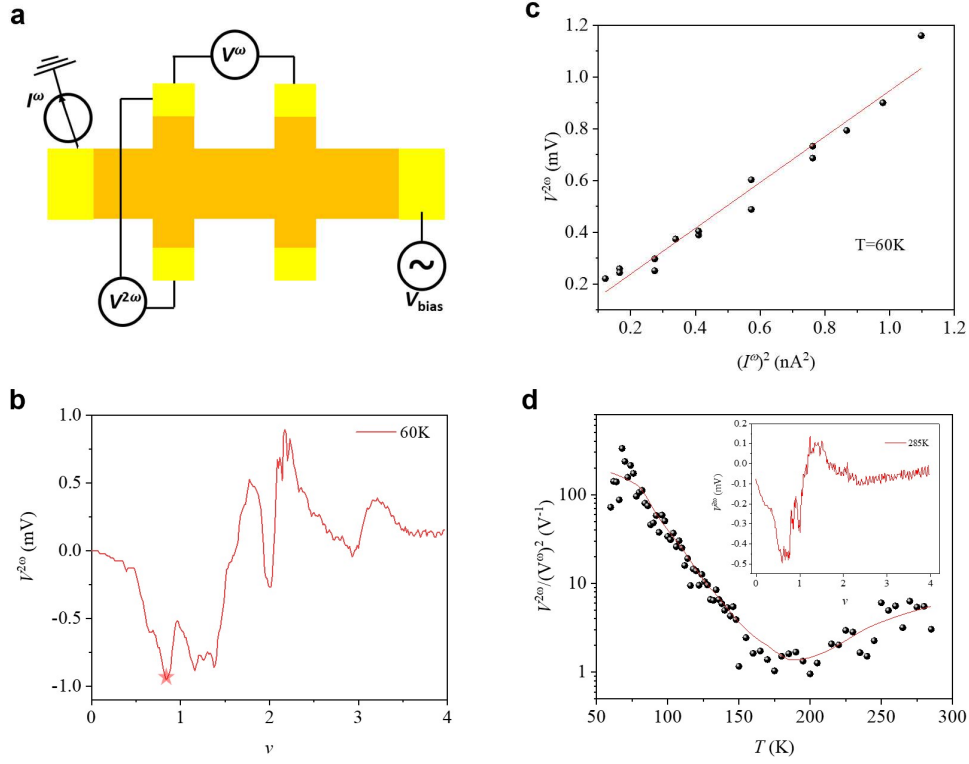

**Fig. 3 | Giant non-linear Hall responses up to room-temperature in twisted bilayer MoS$_2$. a**, Schematic diagram of non-linear Hall measurements. When applying an alternating current $I^\omega$ with frequency $\omega$, transverse non-linear Hall voltage $V^{2\omega}$ at frequency $2\omega$, as well as linear longitudinal voltage $V^\omega$, is detected. **b**, Non-linear Hall voltage $V^{2\omega}$ as a function of moiré filling factors for the 57° twisted bilayer MoS$_2$ device at a temperature of 60 K. $V^{2\omega}$ shows strong response near $v = 1$. **c**, The linear dependence of $V^{2\omega}$ with the square of applied current $(I^\omega)^2$ at $T = 60$ K for filling factor highlighted in **b**. **d**, Temperature dependence of non-linear Hall generation efficiency $\eta = V^{2\omega}/(V^\omega)^2$ for filling factor marked in **b**. Inset: $V^{2\omega}$ versus moiré filling factors at 285 K.

## Conclusions

We demonstrate the transport evidence of room-temperature correlated electronic states and layer-hybridized SU(4) Hubbard model in twisted MoS$_2$ homo-bilayer moiré superlattices with twist angles ~57.5°. Correlated insulating states are clearly observed at moiré band filling factors $v = 1, 2, 3$, providing the evidences of layer-hybridized SU(4) correlated Hubbard model. Moreover, electron-electron correlations in twisted bilayer MoS$_2$ are ultra-strong. The correlated insulating state at $v = 1$ can persist up to a record-high temperature of over 285 K and host a record-large Mott-Hubbard gap of about 126 meV. Additionally, large non-linear Hall responses up to room-temperature are also uncovered in twisted bilayer MoS$_2$, demonstrating the quantum geometrical property of electron wavefunctions. Our results imply that the TMD homo-bilayer moiré superlattices with resonantly enhanced interlayer hybridization favor as a room-temperature correlated quantum system, possessing promising prospect for a rich phase diagram of high-temperature correlated physics and showing the great potential for future technological applications.

**Methods**

**Epitaxial growth of MoS$_2$ single crystal.** High quality single crystal MoS$_2$ was grown on sapphire substrate by van der Waals epitaxial technique in our home-built multisource chemical vapor deposition system with three temperature zones[41]. Sulfur (Alfa Aesar, 99.9%, 15 g) was loaded into zone-I and carried by Ar. MoO$_3$ (Alfa Aesar, 99.999%, 30 mg) was loaded into zone-II and carried by Ar/O$_2$. Single side polished, c-plane (0001) sapphire wafer was loaded in zone-III as the growth substrates. During the van der Waals epitaxial growth of MoS$_2$ on sapphire, the temperatures for the S source, MoO$_3$ source, and sapphire substrate are 120, 530, and 930 °C, respectively. The obtained MoS$_2$ single crystal has high quality with largest domain sizes of ~1 mm.

**Device fabrications.** The as-grown high quality single crystal MoS$_2$ on sapphire substrate was first transferred to SiO$_2$ (around 300 nm)/Si substrate employing ultra-pure water assisted method[42]. The transferred sample is atomically flat on a large scale (>100 μm), confirmed by atomic force microscopy. *h*-BN and few layers graphene were mechanically exfoliated from bulk crystals on 300 nm SiO$_2$, then annealed in the forming gas (150 sccm Ar/10 sccm H$_2$) at 450 °C to clean the residue or contaminations on the surface. Then, the twisted bilayer MoS$_2$ were realized by dry stamping with a modified 'cut and stack' technique based on the transferred MoS$_2$ single crystal on SiO$_2$/Si. A thin polypropylene carbonate (PC) film supported by polydimethylsiloxane (PDMS) on glass slide was used to pick up *h*-BN (15-35 nm) and the first part of MoS$_2$ in sequence with accurate alignment in our home-made transfer station. The second part of MoS$_2$ was rotated by a target angle (accuracy of 0.016°) and subsequently picked up by the first half that is already in contact with the *h*-BN. After picking up several graphene layers as contacts, the whole stacked structure was then dropped on another *h*-BN on SiO$_2$/Si substrate at 160 °C. The PC was removed by chloroform at room temperature. We further patterned the stacks with PMMA resist and CHF$_3$/O$_2$ plasma to define Hall bar structure and expose the edges of graphene by the standard micro-fabrication processes including e-beam lithography (EBL) and reactive ion etching (RIE). Finally, we made one dimensional contact to graphene with Cr/Au (3nm/30nm) and top gate with Ti/Au (3nm/30nm).

**Transport measurements.** Transport measurements were performed in a cryostat with temperature from 2 K to 285 K. We applied standard lock-in techniques with SR830

(built-in resistance value is 10 MΩ) to measure the resistance with an excitation bias of 5 mV-50 mV at the frequency of 30.89 Hz. For non-linear measurement, an excitation voltage up to 0.3 V was applied to detect the relatively weak signature for four terminal measurements, and lock-in amplifiers were set to 0° and 90° for first order current and second order non-linear Hall voltage, respectively.

**Band structure calculations.** The present calculations for the twisted bilayer $MoS_2$ were done within density functional theory using the Vienna ab initio software package[43]. The exchange correlation functionals of Perdew Burke and Ernzerhof (PBE) were used[44], in conjunction with Tkatchenko-Scheffler (TS)[45,46] vdW corrections, which has been shown to give results well consistent with the experimental observations in our previous work on TMD-based moiré superlattices[17]. An energy cutoff of 400 eV for the plane wave basis sets and the Γ-centered k-meshes of 1×1×1 was used for geometry optimization and electronic structure calculations. A vacuum thickness larger than 15 Å was used to avoid artificial interactions between periodic slab images. All atoms were fully relaxed with residual force per atom less than 0.01 eV Å$^{-1}$. Considering the computational cost of superlattice calculations, while the internal atomic positions were fully optimized, the lattice constant for moiré supercell was fixed to a value such that it corresponds to the experimental lattice constant for a 1×1 unit cell.

**Acknowledgements**
We thank Xiaobo Lu for fruitful discussions. We acknowledge supports from the National Key Research and Development Program of China (Grant Nos. 2021YFA1202900, 2020YFA0309600), National Science Foundation of China (NSFC, Grant Nos. 61888102, 11834017, 1207441), the Strategic Priority Research Program of CAS (Grant Nos. XDB30000000& XDB33000000) and the Key-Area Research and Development Program of Guangdong Province (Grant No. 2020B0101340001). K.W. and T.T. acknowledge support from the Elemental Strategy Initiative conducted by the MEXT, Japan (Grant No. JPMXP0112101001), JSPS KAKENHI (Grant Nos. 19H05790, 20H00354 and 21H05233) and A3 Foresight by JSPS.

**Competing interests**
The authors declare no competing interests.

energy contributions to binding in molecules and solids. *J. Chem. Phys.* **132**, 234109 (2010).